\documentclass[preprint,floats,aps,epsfig,nofootinbib,amssymb]{revtex4}
\usepackage{subfigure}
\usepackage{graphicx,epsfig}
\begin{document}

\def\etmiss{E\!\!\!\!\slash_{T}}
\def\ptmiss{p\!\!\!\slash_{T}}
\def\dm{\Delta M_{TA} }
\def\tev{\rm TeV}
\def\gev{\rm GeV}
\def\mev{\rm MeV}
\def\ev{\rm eV}
\def\fbi{\rm fb^{-1}}
\def\nul{\nu_L^{}}
\def\nur{\nu_R^{}}
\def\pslash{\not{\hbox{\kern-4pt $p$}}}
\def\qslash{\not{\hbox{\kern-4pt $q$}}}
\def\lv{\not{\hbox{\kern-4pt $L$}}}

\def\beq{\begin{equation}}
\def\eeq{\end{equation}}
\def\bea{\begin{eqnarray}}
\def\eea{\end{eqnarray}}
\def\nur{\nonumber}
\draft
\preprint{MADPH--08--1509,~~~
NSF--KITP--08--55}
%
%

\title{\Large{The  ``Top Priority" at the LHC}}

\vskip 1cm

\author{Tao Han\footnote{email: {\tt than@hep.wisc.edu} }}
\address{Department of Physics, University of Wisconsin, Madison, WI 53706 \\
KITP, University of California, Santa Barbara, CA 93107}

\date{\today}

\vskip 1cm

\begin{abstract}
The LHC will be a top-quark factory. With 80 million pairs of  top quarks and an additional
34 million 
single tops  produced annually at the designed high luminosity, the properties of this particle 
will be studied to a great accuracy. The fact that the top quark is the heaviest
elementary particle in the Standard Model with a mass right at the electroweak 
scale makes it  tempting to contemplate its role in electroweak
symmetry breaking, as well as its potential as a window to unknown new physics 
at the TeV scale. We summarize the expectations for  
top-quark physics at the LHC, and outline new physics scenarios in which
the top quark is  crucially involved.

\vskip 1in
\hfill

\noindent
To be published as a chapter in the book of 
``{\it Perspectives on the LHC}",  edited by G.~Kane and A.~Pierce, by
World Scientific Publishing Co., 2008.

\end{abstract}

\maketitle

\section{Brief Introduction}

The top quark plays a special role in the Standard Model (SM) and holds
great promise in revealing the secret of new physics beyond the SM.
The theoretical considerations include the following:
\begin{itemize}
\item With the largest Yukawa coupling $y_t\sim 1$ among the SM fermions, 
and a mass at the electroweak scale $m_t\sim v/\sqrt 2$ (the vacuum
expectation value of the Higgs field),  the top quark is naturally related to 
electroweak symmetry breaking (EWSB), and may reveal new strong dynamics \cite{Hill:2002ap}.
\item The largest contribution to the quadratic divergence of the SM Higgs mass comes
from the top-quark loop, which implies the immediate need for new physics at the Terascale
for a natural EW theory \cite{Giudice:2008bi},
with SUSY and Little Higgs as prominent examples. 
\item Its heavy mass opens up a larger phase space for its decay to heavy
states $Wb,\ Zq,\ H^{0,\pm}q$, {\it etc.}
\item Its prompt decay much shorter than the  QCD scale offers the opportunity to explore
the properties of  a ``bare quark", such as its spin, mass, and couplings.
\end{itemize}

Top quarks will be copiously produced at the LHC. The production and decay are
well understood in the SM. Therefore, detailed studies of the top-quark physics can
be rewarding for both testing the SM and searching for new physics \cite{Quadt:2006jk}.

\section{Top Quark in The Standard Model}\label{sec1.2}

In the SM, the top quark and its interactions can be described by
\bea
-{\cal L}_{SM} &=&  m_t \bar t t + {m_t\over v} H \bar t t 
+ g_s \bar t \gamma^\mu T^a t G_\mu^a + e Q_t \bar t \gamma^\mu t A_\mu \\
&+& {g\over \cos\theta_w} \bar t  \gamma^\mu  (g_V+g_A\gamma^5) t Z_\mu 
+ {g\over \sqrt 2} \sum_q^{d,s,b} V_{tq} \bar t \gamma^\mu P_L q W^-_\mu +h.c.\ \ \ \ 
\nonumber
\eea
Besides the well-determined gauge couplings at the electroweak scale,
the other measured parameters of the top quark  are listed in Table \ref{tab:I}.
\begin{table}[h]
\caption{Experimental values for the top quark parameters \cite{pdg}.}
{\begin{tabular}{@{}cccc@{}} \toprule
$m_t$ (pole)  & $|V_{tb}|$ & $|V_{ts}|$ & $|V_{td}|$ \\
 \colrule
(172.7 $\pm$ 2.8) GeV~~~ & $>0.78$~~~ & $(40.6\pm 2.6)\times 10^{-3}$~~~ & 
$(7.4\pm 0.8)\times 10^{-3}$
\\ \botrule
\end{tabular}
}
\label{tab:I}
\end{table}

The large top-quark mass is important
since it  contributes significantly to the electroweak radiative corrections. For instance, 
the one-loop corrections to the electroweak gauge boson mass can be cast  in the form 
\beq
\Delta r = -{3G_F m_t^2\over 8\sqrt 2 \pi^2\tan^2\theta_W} +
{3G_F M_W^2\over 8\sqrt 2 \pi^2}\left(\ln{m_H^2\over M_Z^2} -{5\over 6}\right).
\eeq
With the $m_t$ value in Table \ref{tab:I}, the best global fit in the SM yields a
Higgs mass $m_H=89^{+38}_{-28}$ GeV \cite{pdg}. 
The recent combined result from CDF and D0 at the Tevatron Run II gave the new
value \cite{Brubaker:2006xn}
\beq
m_t = 171.4 \pm 2.1 \ {\rm GeV}.
\eeq
The expected accuracy  of $m_t$ measurement at the LHC is better than 
1 GeV \cite{Etienvre:2006ph}, with errors dominated by the systematics. 

To directly determine the left-handed $V$-$A$ gauge coupling of the top quark
in the weak charged current, leptonic angular distributions and $W$ polarization 
information would be needed \cite{gordy}.
No direct measurements are available yet for the electroweak neutral current couplings,
$g_V^t = T_3/2 -Q_t \sin^2\theta_W,\ g_A^t = -T_3/2$ and $Q_t=+2/3$,
although there are proposals to study them via the associated production processes
$t\bar t \gamma,\ t\bar t  Z$ \cite{Baur:2001si}.
The indirect global fits however indicate the consistency with these SM 
predictions  \cite{pdg}.

\subsection{Top-Quark Decay in the SM}

Due to the absence of the flavor-changing neutral currents at tree level
in the SM (the Glashow-Iliopoulos-Maiani mechanism), 
the dominant decay channels for a top quark will be through the weak charged-currents, 
with the partial width given by \cite{twidth}
\beq
\Gamma(t\to W^+q) = {|V_{tq}|^2  m_t^3\over 16\pi v^2} (1-r_W)^2 (1+2r_W)
\left[1-{2\alpha_s\over 3\pi}({2\pi^2\over 3}-{5\over 2}) \right],
\label{eq:width}
\eeq
where $r_W = M_W^2/m_t^2$. The subsequent decay of $W$ to the final state
leptons and light quarks is well understood. Two important features are noted:
\begin{itemize}
\item
Since $|V_{tb}| \gg |V_{td}|, |V_{ts}|$, a top quark
will predominantly decay into a $b$ quark. 
While $V_{ts},\ V_{td}$ may not be practically measured via the top-decay processes,
effective $b$-tagging at the Tevatron experiments has served to put a bound on the ratio
\beq
{B(t\to Wb) \over B(t\to Wq)} = { |V_{tb}|^2 \over { |V_{td}|^2+ |V_{ts}|^2 + |V_{tb}|^2 } }, 
\eeq
that leads to the lower bound for $|V_{tb}|$ in Table  \ref{tab:I}.
\item
Perhaps the most significant aspect
of Eq.~(\ref{eq:width}) is the numerics:
\beq
\Gamma(t\to W^+q) \approx 1.5\ {\gev} \approx {1\over 0.5\times 10^{-24}\ {\rm s}} > 
\Lambda_{QCD}\sim 200\ {\mev}.
\nonumber
\eeq
This implies that a top quark will promptly decay via weak interaction before QCD sets 
in for hadronization \cite{tlife}. 
So no hadronic bound states (such as $\bar t t, \bar tq$, {\it etc.})
would be observed. The properties of a ``bare quark" may be accessible
for scrutiny.
\end{itemize}

It is interesting to note that in the top-quark rest frame, the longitudinal polarization
of the $W$ is the dominant mode. The ratio between the two available modes is
\beq
{\Gamma(t\to b_L\ W_{\lambda=0})\over \Gamma(t\to b_L\ W_{\lambda=-1})} 
= {m_t^2\over 2M_W^2}.
\label{eq:pol}
\eeq

\subsection{Top-Quark Production in the SM}

\subsubsection{$t\bar t$ production via QCD}

Historically, quarks were discovered via their hadronic bound states, most notably
for the charm quark via $J/\psi(\bar c  c)$ and bottom quark via $\Upsilon(\bar b b)$. 
Due to the prompt decay of the top quark, its production mechanisms and search
strategy are quite different from the traditional one.

\begin{figure}[t]
\centerline{\psfig{file=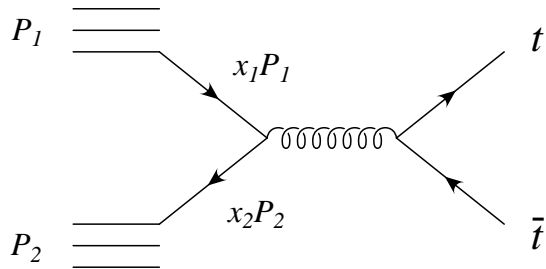,width=7.8cm}}
\caption{Top-quark pair production in hadronic collisions via QCD interaction.
This figure is taken from Ref.~\cite{Willenbrock:2002ta}. }
\label{fig:pptt}
\end{figure}

The leading processes are the open flavor pair production from the QCD strong 
interaction, as depicted in Fig.~\ref{fig:pptt}. The contributing subprocesses are from 
\beq
q\bar q,\ gg \to t\bar t.
\label{tt-TT}
\eeq
The cross sections have been calculated rather reliably to the next-to-leading
order \cite{Nason:1987xz} 
and including the threshold resummations \cite{Laenen:1993xr,Bonciani:1998vc},
as given in Table \ref{tab:xsectiontt}.
\begin{table}[h]
\caption{
Cross sections, at next-to-leading-order in QCD, for top-quark
production via the strong interaction at the Tevatron and the LHC
\cite{Bonciani:1998vc}.  Also shown is the percentage of the total cross
section from the quark-antiquark-annihilation and gluon-fusion subprocesses.}
{\begin{tabular}{@{}c|c|c|c@{}} 
\toprule
&$\sigma_{\rm NLO}$ (pb)&$q\bar q\to t\bar t$&$gg\to t\bar t$ \\
\toprule
Tevatron ($\sqrt s=1.8$ TeV $p\bar p$)&$4.87\pm 10\%$&$90\%$&$10\%$ \\
\toprule
Tevatron ($\sqrt s=2.0$ TeV $p\bar p$)&$6.70\pm 10\%$&$85\%$&$15\%$ \\
 \colrule
LHC ($\sqrt s=14$ TeV $pp$)&$803\pm 15\%$&$10\%$&$90\%$\\ 
\hline 
\end{tabular}
}
\label{tab:xsectiontt}
\end{table}

\begin{figure}[tb]
\centerline{\psfig{file=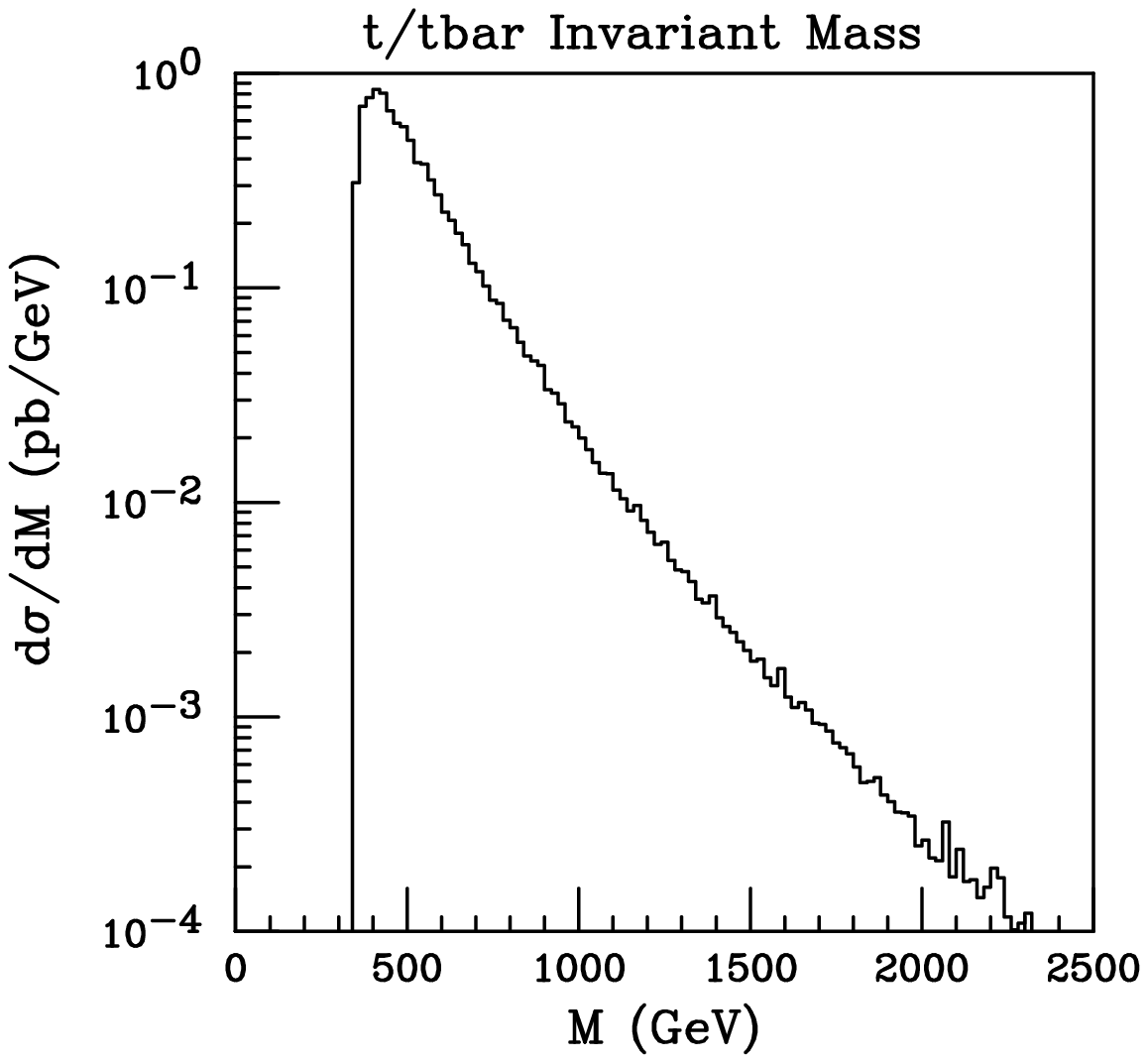,width=7.8cm},\psfig{file=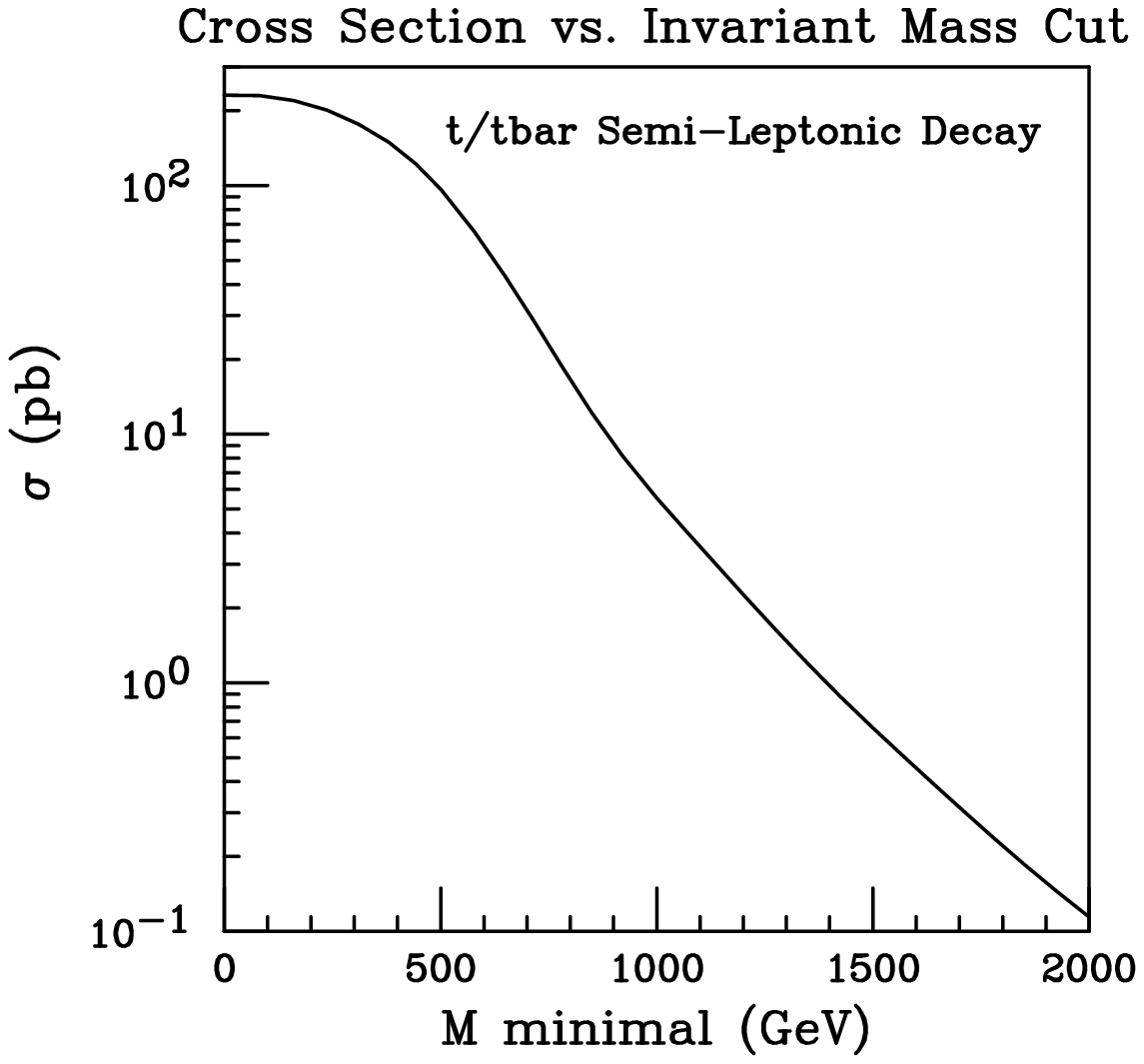,width=7.8cm}}
\caption{
(a) Invariant mass distribution of $t\bar t$ at the LHC and (b) integrated
cross section versus a minimal cutoff on  $m(t\bar t)$. 
Decay branching fractions of one top decaying
hadronically and the other leptonically ($e,\mu$) have been included. }
\label{fig:mtt}
\end{figure}

\noindent
Largely due to the substantial gluon luminosity at higher energies, 
the $t\bar t$ production rate is increased by more than a factor of 100
from the Tevatron to the LHC. Assuming an annual luminosity at the LHC of
$10^{34}$ cm$^{-2}$ s$^{-1} \Rightarrow 100$ fb$^{-1}/$year, one expects
to have 80 million top pairs produced. It is truly a ``top factory".
In Fig.~\ref{fig:mtt}(a), we plot the invariant mass distribution, which is
important to understand when searching for new physics in the $t\bar t$
channel. Although the majority of the events are produced near the threshold
$m(t\bar t) \sim 2m_t$, there is still a substantial cross section even above 
$m(t\bar t)\sim$ 1 TeV, about 5 pb. This is illustrated in Fig.~\ref{fig:mtt}(b),
where the integrated cross section is given versus a minimal
cutoff on $m(t\bar t)$ and decay branching fractions of one top decaying
hadronically and the other leptonically have been included.

It should be noted that the forward-backward charge asymmetry of the
$t\bar t$ events can be generated by higher order corrections, reaching
$10-15\%$ at the partonic level from QCD \cite{Kuhn:1998jr} and $1\%$ from the 
electroweak \cite{Bernreuther:2005is}.

\subsubsection{Single top production via weak interaction}

\begin{figure}[tb]
\centerline{\psfig{file=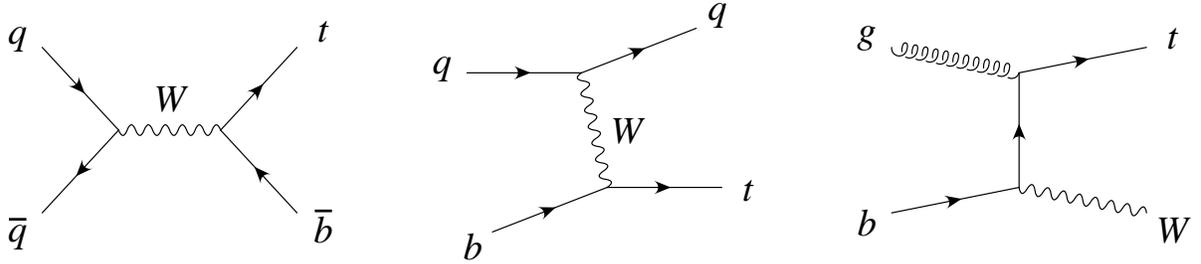,width=16cm}}
\caption{Single top-quark production in hadronic collisions via the 
charged-current weak interaction.
This figure is taken from Ref.~\cite{Willenbrock:2002ta}. }
\label{fig:ppt}
\end{figure}

As discussed in the last section, the charged-current weak interaction is
responsible for the rapid decay of the top quark. In fact, it also participates
significantly in the production of the top quark as well \cite{Willenbrock:cr}.
The three classes of
production processes, $s$-channel Drell-Yan, $t$-channel $Wb$ fusion,
and associated $Wt$ diagrams, are plotted in Fig.~\ref{fig:ppt}. 
Two remarks are in order:
\begin{itemize}
\item The single top production is proportional to the quark mixing element
$|V_{tb}|^2$ and thus provides the direct measurement for it,
currently \cite{Abazov:2006gd} $0.68 < |V_{tb}| \le 1$ at the $95\%$ C.L. 
\item The $s$-channel and $t$-channel can be complementary in the search
for new physics such as a $W'$ exchange \cite{Cao:2007ea}.
\end{itemize}

For the production 
rates \cite{Smith:1996ij,Stelzer:1997ns,Zhu:uj,Kidonakis:2006bu,Kidonakis:2007ej}, 
the largest of
all is the $t$-channel $Wb$ fusion. It is nearly one third of the QCD production
of the $t\bar t$ pair. Once again, it is mainly from the enhancement of the
longitudinally polarized $W$. The total cross sections for these processes
at Tevatron \cite{Kidonakis:2006bu} and LHC energies \cite{Kidonakis:2007ej}
are listed in Table \ref{tab:xsectiont} \cite{Smith:1996ij,Stelzer:1997ns,Zhu:uj}.
We see the typical change of the production rate from the Tevatron to the
LHC: A valence-induced process (DY-type) is increased by about an order
of magnitude; while the gluon- or $b$-induced processes are enhanced by
about a factor of 100.

\begin{table}[h]
\caption{
Cross sections, at next-to-leading-order in QCD, for top-quark
production via the charged current weak interaction at the Tevatron and the LHC.}
{\begin{tabular}{@{}c|c|c|c@{}} 
\toprule
$\sigma({\rm pb}) $ &$s$-channel & $t$-channel & $Wt$ \\
\hline
Tevatron ($\sqrt s=2.0$ TeV $p\bar p$)&$0.90\pm 5\%$&$2.1\pm 5\%$&$0.1\pm 10\%$ \\
\hline
LHC ($\sqrt s=14$ TeV $pp$)&$10.6\pm 5\%$&$250\pm 5\%$&$75\pm 10\%$\\
\hline 
\end{tabular} }
\label{tab:xsectiont}
\end{table}

\subsubsection{Top quark and Higgs associated production}
Of fundamental importance is the measurement of the top-quark Yukawa coupling.
The direct probe to it at the LHC is via the processes \cite{Marciano:1991qq}
\beq
q\bar q,\  gg \to t\bar t H.
\eeq
The cross section has been calculated to the next-to-leading-order (NLO)  in 
QCD \cite{Beenakker:2001rj,Dawson:2003zu} 
and the numerics are given in Table \ref{tab:H}.
The cross section ranges are estimated from the uncertainty of the QCD
scale. 
\begin{table}[h]
\caption{Total cross section at the NLO in QCD for top-quark and Higgs associated
production at the LHC \cite{Dawson:2003zu}. }
{\begin{tabular}{@{}c|c|c|c@{}} 
\toprule
$m_H$ (GeV) & 120 & 150 & 180 \\
\hline
$\sigma$ (fb)  &  634$-$719 & 334$-$381& 194$-$222 \\
\hline 
\end{tabular} }
\label{tab:H}
\end{table}

The production rate at the LHC seems quite feasible for the signal observation. 
It was claimed \cite{Desch:2004kf} that a $15\%$ accuracy for 
the Yukawa coupling measurement may be achievable with a luminosity
of 300 fb$^{-1}$. Indeed, the decay channel $H\to \gamma\gamma$ should
be useful for the search and study in the mass range of 
$100 <m_H <150$ GeV \cite{unknown:1999fr,Zhou:1993at}.
However, the potentially large backgounds and the complex
event topology, in particular the demand on the detector performance,
make the study of the leading decay $H\to b \bar b$
very challenging \cite{Benedetti:2007sn}.

\section{New Physics in Top-quark Decay}

The high production rate for the top quarks at the LHC provides a great
opportunity to seek out top-quark rare decays and search for new physics
Beyond the Standard Model (BSM).  Given the annual yield of 80 million
$t\bar t$ events plus  $34$ million single-top events, 
one may hope to search for rare decays
with a branching fraction as small as $10^{-6}$.

\subsection{Charged Current Decay: BSM}
The most prominent examples for top-quark decay beyond the SM via charged-currents
may be the charged Higgs in SUSY or with an extended Higgs sector,
and charged technicolor particles
\beq
t \to H^+ b,\ \ \pi^+_T b.
\eeq
Experimental searches have been conducted at the Tevatron \cite{Abazov:2001md},
and some simulations are performed for the LHC as 
well \cite{Hashemi:2006qg,Quadt:2006jk}.
It is obvious that as long as those channels are kinematically accessible
and have a sizable branching fraction, the observation should be straightforward.
In fact, the top decay to a charged Higgs may well be the leading channel for
$H^\pm$ production. 

More subtle new physics scenarios may not show up with the above easy
signals. It may be desirable to take a phenomenological
approach to parameterize the top-quark interactions beyond the 
SM \cite{gordy,Tait:2000sh},
and experimentally search for the deviations from the SM.
Those ``anomalous couplings" can be determined in a given theoretical
framework, either from loop-induced processes or from a new flavor structure.
One can write the interaction terms as
\bea
{\cal L}_{CC} = 
 {g\over \sqrt 2} \left(\  \bar t (1+\delta_L) \gamma^\mu P_L q W^-_\mu + 
  \bar t \delta_R  \gamma^\mu P_R q W^-_\mu \right)+h.c.
\eea
The expected accuracy of the measurements on $\delta_{L,R}$ 
is about $1\%$ \cite{Tait:2000sh,Quadt:2006jk},  thus testing the top-quark 
chiral coupling. 

\subsection{Neutral Current Decay: BSM}
Although there are no Flavor-Changing Neutral Currents (FCNC) at tree level in the SM, 
theories beyond the SM
quite often have new flavor structure, most notably for SUSY and technicolor
models. New symmetries or some alignment mechanisms will have to be
utilized in order to avoid excessive FCNC. It is  nevertheless prudent to keep in mind
the possible new decay modes of the top quark such as the SUSY decay channel
\beq
t \to \tilde t \tilde\chi^0.
\eeq

Generically, FCNCs can always be generated  at loop level. It has been shown
that the interesting decay modes
\beq
t\to Zc,\ \ Hc,\ \  \gamma c,\ \  gc
\eeq
are highly suppressed  \cite{Eilam:1990zc,Cao:2007dk}
with branching fractions typically
 $10^{-13} - 10^{-10}$  in the SM, and  $10^{-7} - 10^{-5}$ in the MSSM. 
It has been shown that the branching fractions can be enhanced significantly 
in theories beyond the SM and MSSM, reaching above 
 $10^{-5}$ and even as high as $1\%$ \cite{AguilarSaavedra:2004wm}.

One may again take the effective operator approach to parameterize the interactions.
After the electroweak symmetry breaking, one can write 
them as \cite{Peccei:1989kr,Han:1998tp,Han:1996ep}
\bea
{\cal L}_{NC} &= &
 {g\over  2 \cos\theta_w} \sum_{\tau=\pm,q=c,u}  \kappa_\tau \bar t  \gamma^\mu 
 P_\tau  q Z_\mu + h.c. \\
& +& g_s \sum_{q=c,u}{\kappa^g_q \over \Lambda} 
 \bar t \sigma^{\mu\nu} T^a t G_{\mu\nu}^a +
 e Q_t \sum_{q=c,u}{\kappa^\gamma_q \over \Lambda}  
 \bar t \sigma^{\mu\nu} t A_{\mu \nu} + h.c.
\eea

The sensitivities for the anomalous couplings
have been studied  at the LHC by the ATLAS  Collaboration \cite{Carvalho:2007yi},
as listed in Table \ref{tab:fcnc}
\begin{table}[h]
\caption{$95\%$ C.L.~sensitivity of the branching fractions for the top-quark decays
via FCNC couplings at the LHC \cite{Carvalho:2007yi}.
}
{\begin{tabular}{@{}c|c|c@{}} 
\toprule
Channel  & 10 $\fbi$ & 100 $\fbi$ \\
\hline
$t\to Z q$   &  $3.1\times 10^{-4}$  & $6.1\times 10^{-5}$  \\
\hline 
$t\to \gamma q$   &  $4.1\times 10^{-5}$  & $1.2\times 10^{-5}$  \\
\hline 
$t\to g q$   &  $1.3\times 10^{-3}$  & $4.2\times 10^{-4}$  \\
\hline 
\end{tabular} }
\label{tab:fcnc}
\end{table}

\section{Top Quarks in Resonant Production}

The most striking signal of new physics in the top-quark sector 
is the resonant production via a heavy intermediate state $X$.
With some proper treatment to identify the top decay products, 
it is possible to reconstruct the resonant kinematics.
One may thus envision fully exploring its properties in the c.m.~frame.

\subsection{$X \to t \bar t, \ t\bar b$}\label{sec:xtt}

Immediate examples of the resonant states include Higgs bosons \cite{He:1998ie},
new gauge bosons  \cite{Agashe:2007ki},  
Kaluza-Klein excitations of gluons \cite{Lillie:2007ve}  and 
gravitons \cite{Fitzpatrick:2007qr},
Technicolor-like dynamical states \cite{Hill:2002ap,Quadt:2006jk,Choudhury:2007ux} etc. 

\begin{figure}[tb]
\centerline{\psfig{file=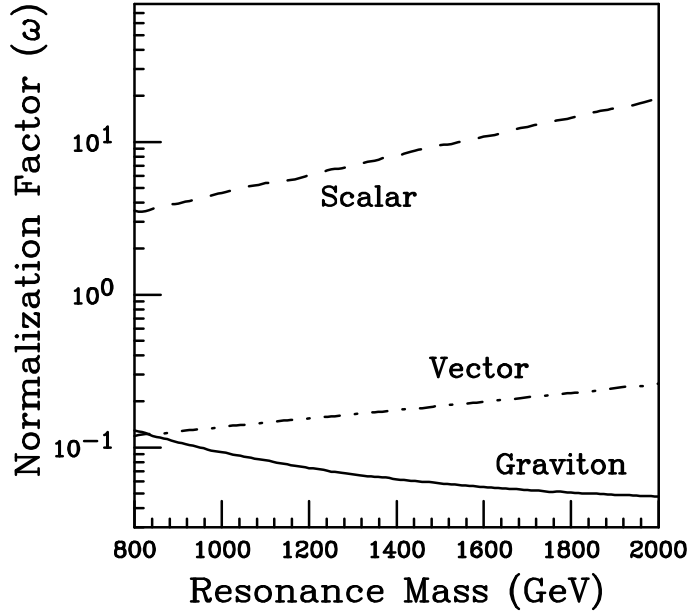,width=9cm}}
\caption{Normalization factor versus the resonance mass 
for the scalar (dashed) with a width-mass ratio of $20\%$, 
vector (dot-dashed) with 5\%,  and graviton (solid) 2\%, respectively.  
The region above each curve represents values of $\omega$ that give 5$\sigma$ 
or greater statistical  significance with 10 fb$^{-1}$ integrated luminosity.}
\label{fig:X}
\end{figure}

The signal can be generically written as 
\bea
\nonumber
\sigma(pp \to X\to  t\bar t) &=& \sum_{ij} \int dx_1 dx_2 f_i(M_X^2, x_1) f_j(M_X^2,x_2) \\
&\times&{ 4\pi^2 (2J+1)\over s} { \Gamma(X\to ij) B(X\to t\bar t) \over M_X}.
\eea
Thus the observation of this class of signals depends on the branching fraction of 
$X\to t\bar t$ as well as its coupling to the initial state partons. 
Figure \ref{fig:X} quantifies the observability for a bosonic resonance (spin 0,1,2)
for a mass up to 2~TeV at the LHC \cite{Barger:2006hm} via $q\bar q, gg \to X \to t\bar t$. 
The vertical axis gives the normalization factors ($\omega$) for the cross section rates
needed to reach a $5\sigma$ signal with a luminosity of 10 fb$^{-1}$.
 The normalization $\omega=1$ defines the benchmark for the spin 0, 1 and 2 resonances.  They correspond to the SM-like Higgs boson, a $Z'$ with electroweak coupling strength and left (L) or right (R) chiral couplings to SM fermions, and the Randall-Sundrum graviton $\tilde h$ with the couplings 
 scaled with a cutoff scale 
 as $\Lambda^{-1}$ for $\tilde h q\bar q$, and $(\Lambda \ln(M^*_{pl}/\Lambda))^{-1}$ for $\tilde h gg$, respectively. We see that a $Z'$ or a graviton should be easy to observe, but a Higgs-like broad 
 scalar will be difficult to identify in the $t\bar t$ channel. 

It is of critical importance to reconstruct the c.m.~frame
of the resonant particle, where the fundamental properties
of the particle can be best studied.  It was demonstrated \cite{Barger:2006hm} 
that with the semi-leptonic decays of the two
top quarks, one can effectively reconstruct the events in the c.m.~frame.
This relies on using the $M_W$ constraint to determine the missing
neutrino momentum, while it is necessary to also make use of $m_t$
to break the two-fold ambiguity for two possible $p_z(\nu)$ solutions. 
Parity and CP asymmetries \cite{Atwood:2000tu} can  be studied.

Top-quark pair events at the high invariant mass are obviously important to 
search for and study new physics. In this new territory there comes a new
complication: When the top quark is very energetic, $\gamma = E/m_t \sim 10$, 
its decay products may be too collimated to be individually resolved by
the detector $-$ recall that the granularity of the hadronic calorimeter at the 
LHC is roughly $\Delta\eta\times \Delta\phi\sim 0.1\times 0.1$.
This is a generic problem relevant to any fast-moving top quarks from
heavy particle decays \cite{Lillie:2007ve,Barger:2006hm,Skiba:2007fw} 
(see the next sections). The interesting questions to be addressed may include: 
\begin{itemize}
\item To what extent can we tell a ``fat top-jet" from a massive QCD jet due to showering?
\item To what extent can we tell a ``fat $W$-jet" from a massive QCD jet?
\item Can we make use of a non-isolated lepton inside the top-jet ($b\ell\nu$)
for the top-quark identification and reconstruction?
\item Can we do $b$-tagging for the highly boosted top events?
\end{itemize}
These practical issues would become critical to understand the events
and thus for new physics searches.
Detailed studies including the detector effects will be needed to reach
quantitative conclusions.

\subsection{$T \to t Z, \ tH,\ bW$}
In many theories beyond the SM, there is a top-quark partner. These are 
commonly motivated by the ``naturalness" argument, the 
need to cancel the quadratic divergence in the Higgs mass radiative correction,
most severely from the top-quark loop. Besides the scalar top quark in SUSY, the
most notable example is the Little Higgs theory \cite{Schmaltz:2005ky}.
If there is no discrete symmetry, the top partner $T$ will decay to SM particles 
in the final state, leading to fully a reconstructable fermionic resonance.

\begin{figure}[tb]
\centerline{\psfig{file=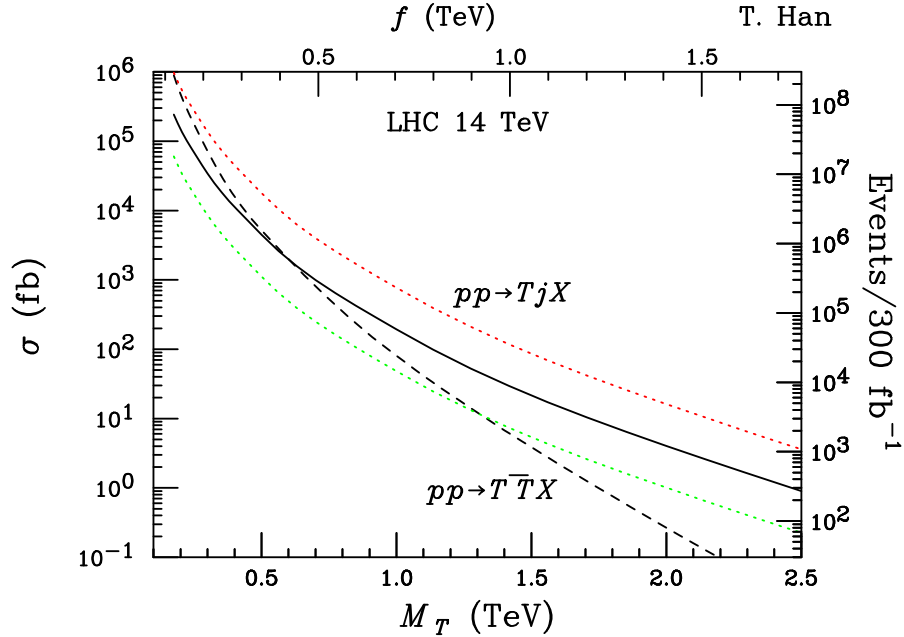,width=12cm}}
\caption{Production of  the top-quark partner $T$ in pair and singly at the
LHC versus its mass. The Yukawa coupling ratio $\lambda_1/\lambda_2$
has been taken to be 2 (upper dotted curve) 1 (solid) and 1/2 (lower dotted),
respectively. The $T\bar T$ pair production via QCD includes an 
NLO $K$-factor (dashed curve).}
\label{fig:Tj}
\end{figure}

\begin{figure}[tb]
\centerline{\psfig{file=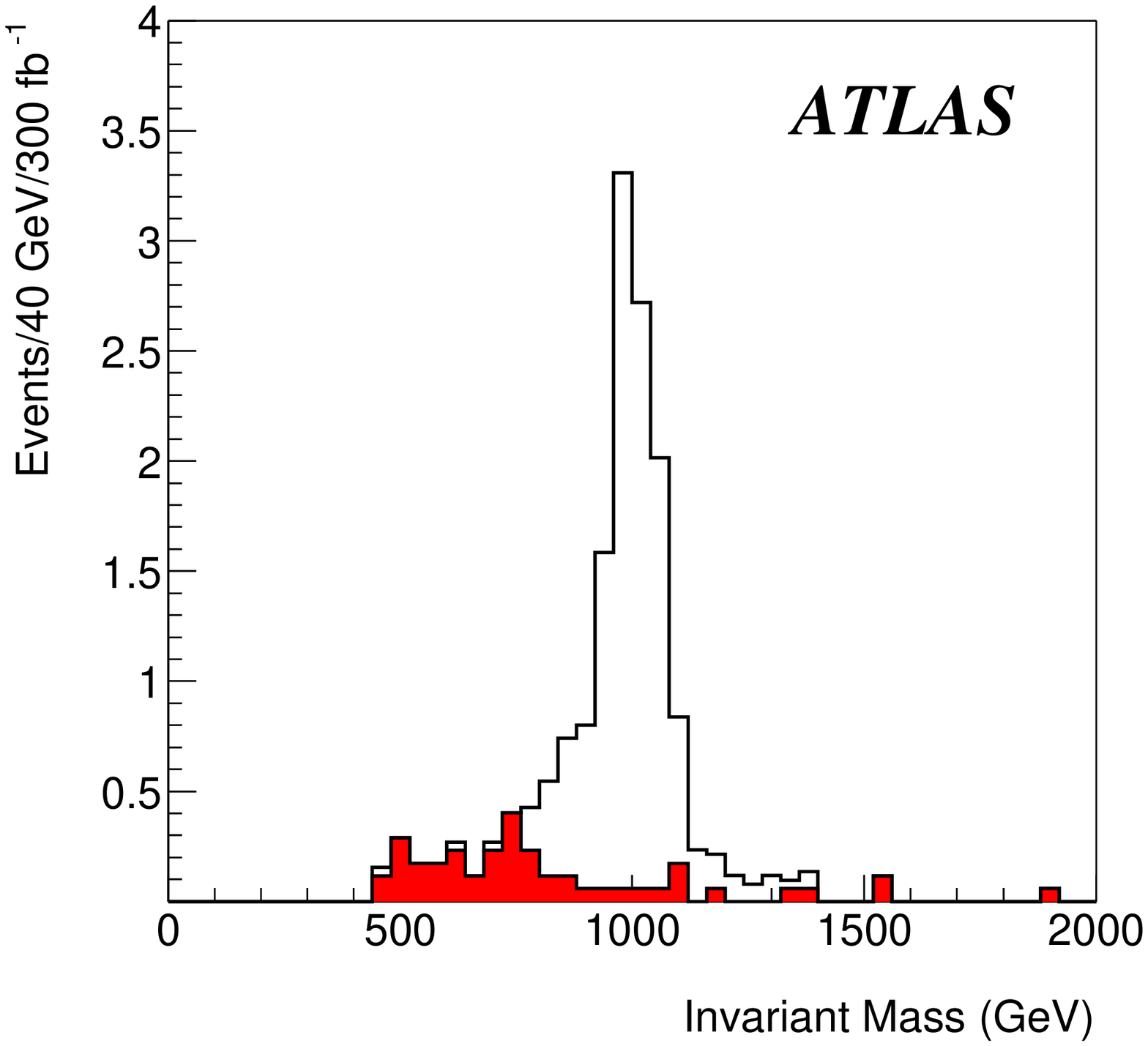,width=8cm}
\psfig{file=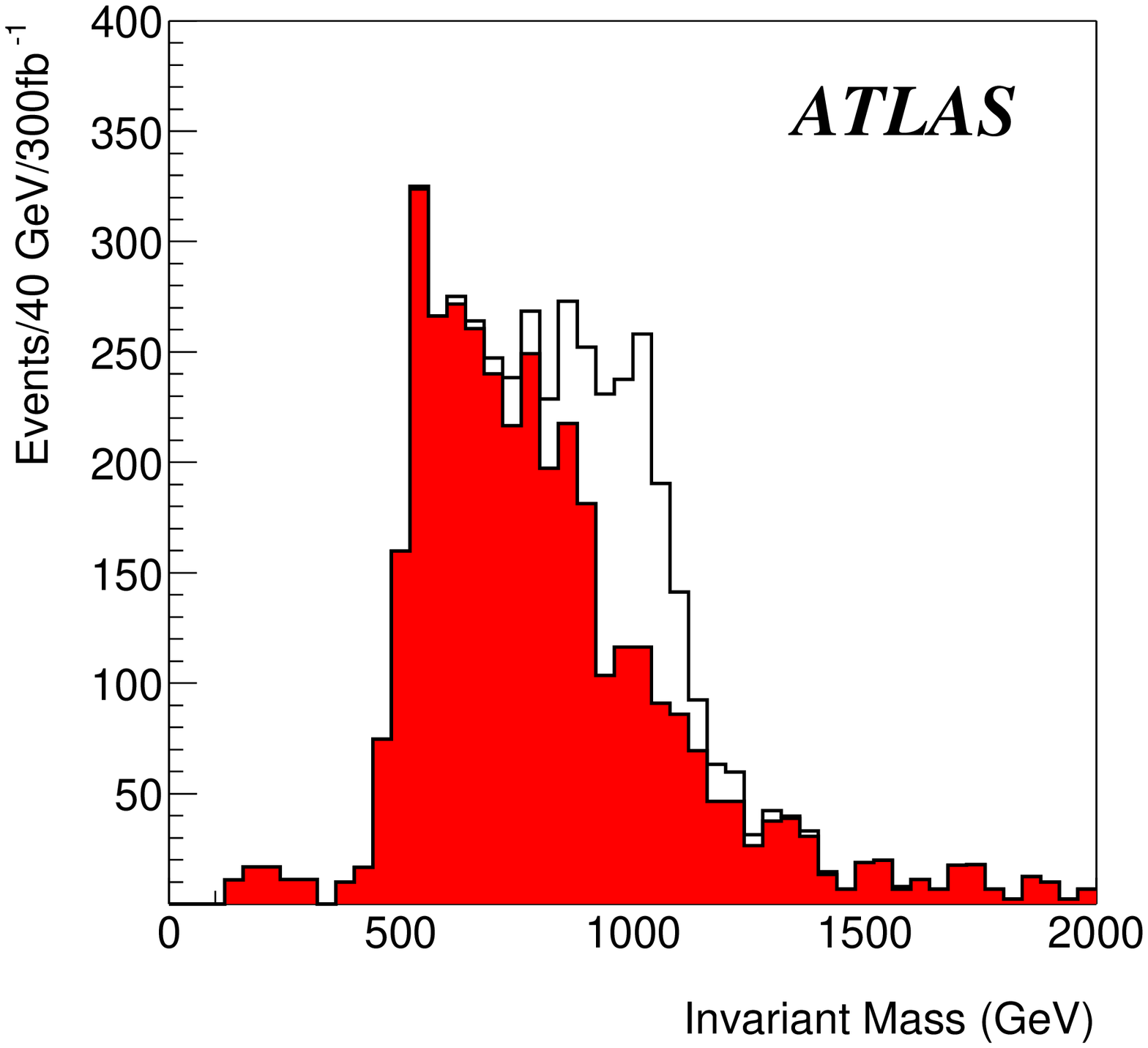,width=8cm}}
\caption{Observability for the decays (a) $T\to tZ$ and (b) $T\to bW$ at the 
ATLAS \cite{Azuelos:2004dm}. }
\label{fig:TZ}
\end{figure}

It was pointed out \cite{Han:2003wu} that the single $T$ production via
the weak charged-current  may surpass the pair production via the QCD
interaction due to the longitudinal gauge boson enhancement for the
former and the phase space suppression for the latter. This is shown in Fig.~\ref{fig:Tj}.
Subsequent simulations \cite{Azuelos:2004dm}
performed by the ATLAS collaboration demonstrated the clear
observability for the signals above the backgrounds at the LHC for $T\to tZ,\ bW$ 
with a mass $M_T = 1$ TeV,  as seen in Fig.~\ref{fig:TZ}.

\section{Top-rich Events for New Physics}

Although the top-quark partner is strongly motivated for a natural electroweak 
theory, it often results in excessively large corrections to the low energy
electroweak observables.
In order to better fit the low energy measurements,
a discrete symmetry is often introduced, such as the R-parity in SUSY,
KK-parity in UED, and T-parity in LH \cite{Cheng:2003ju}.
The immediate consequence for collider phenomenology is the appearance
of a new stable particle that may provide the cold dark matter candidate,
and  results in missing energy in collider experiments.\footnote{Alternatively, the
breaking of the R-parity \cite{Barbier:2004ez} or the T-parity \cite{Hill:2007nz} 
would lead to different collider phenomenology \cite{Barger:2007df}. }

\subsection{$T \bar T$ pair production}

\begin{figure}[tb]
\centerline{\psfig{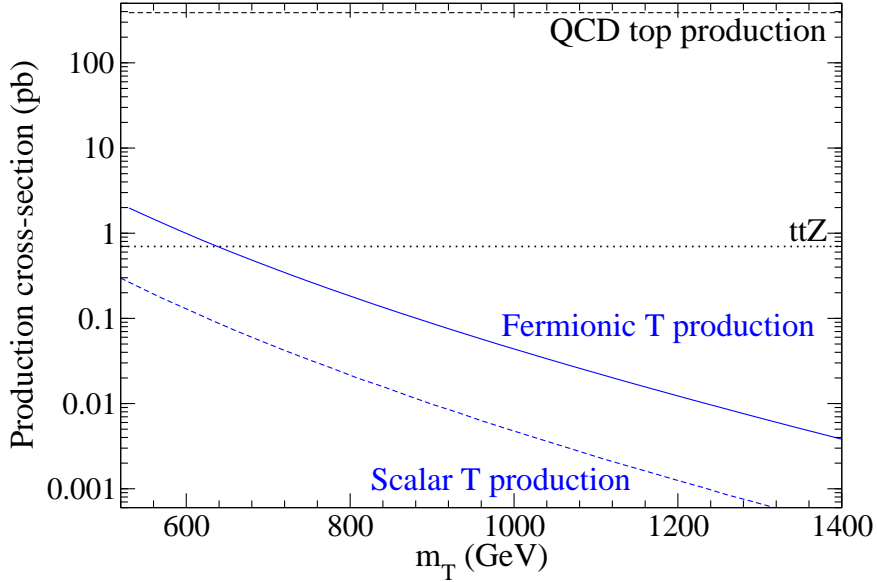}}
\caption{Leading order total cross section for the top partner $T\bar T$ production at the LHC
versus its mass \cite{us}.
Both spin-0 and spin-1/2 top partners are included. 
The QCD $t\bar t$ and the SM $t\bar t Z$ backgrounds
are indicated by the horizontal lines.}
\label{fig:Ts}
\end{figure}

The top partner has similar quantum numbers to the top quark, and thus
is commonly assigned as a color triplet. This leads to their production in QCD
\beq
q\bar q, \ gg \to T \bar T.
\eeq
The production cross section is shown in Fig.~\ref{fig:Ts} for both
 spin-0 and spin-1/2 top partners. Although there is a difference of a factor
 of 8 or so (4 from spin state counting and the rest from threshold effects) 
in the cross sections, it is still challenging to tell a scalar and a fermionic
 partner apart \cite{us,Cheng:2005as,Meade:2006dw} due to the lack of
 definitive features. 
 
Due to the additional discrete symmetry, the top partner cannot decay to a
SM particle alone. Consequently, $T\to tA^0$, leading to $t\bar t$ pair 
production plus large mixing energy. The crucial parameter to characterize 
the kinematical features is the mass difference $\dm = m_T - m_{A}$.
For $\dm \gg m_t$, the top quark as a decay product will be
energetic and qualitatively different from the SM background. But if
$\dm \approx m_t$, then the two will have very little difference,
making the signal difficult to separate out.
Depending on the top-quark decay, we present two classes of signals.

\subsubsection{$t \bar t$ pure hadronic decay}

\begin{figure}[tb]
\centerline{\psfig{file=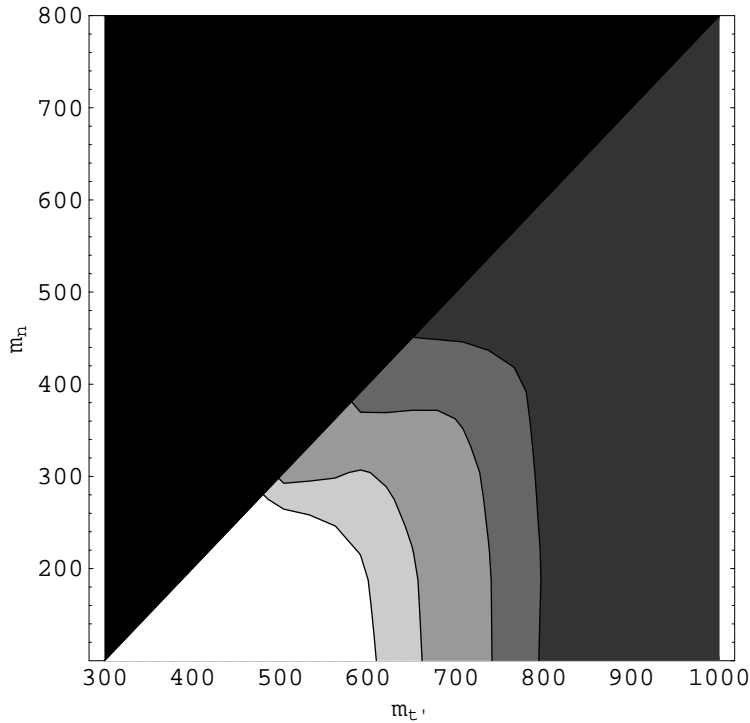,width=10cm} }
\caption{Contour in $m_{\tilde t}-m_N$ for $\tilde t \to tN$
for the statistical significance of a scalar $\tilde t$
at the LHC with an integrated luminosity of 100 fb$^{-1}$.
Purely hadronic decays are considered.}
\label{fig:MR}
\end{figure}

For both $t\bar t$ to decay hadronically \cite{Meade:2006dw,Matsumoto:2006ws},
the signal will be 6 jets plus missing
energy. While it has the largest decay rate, the backgrounds would be
substantial as well. With judicious acceptance cuts, the signal observability
for $\dm > 200$ GeV was established, as seen in Fig.~\ref{fig:MR}.
Possible measurements of the absolute mass scale and its spin of the
top partner were considered \cite{us,Meade:2006dw}, but the determination
remains difficult.

\subsubsection{$t \bar t$ semi-leptonic decay}

\begin{figure}[tb]
\centerline{\psfig{file=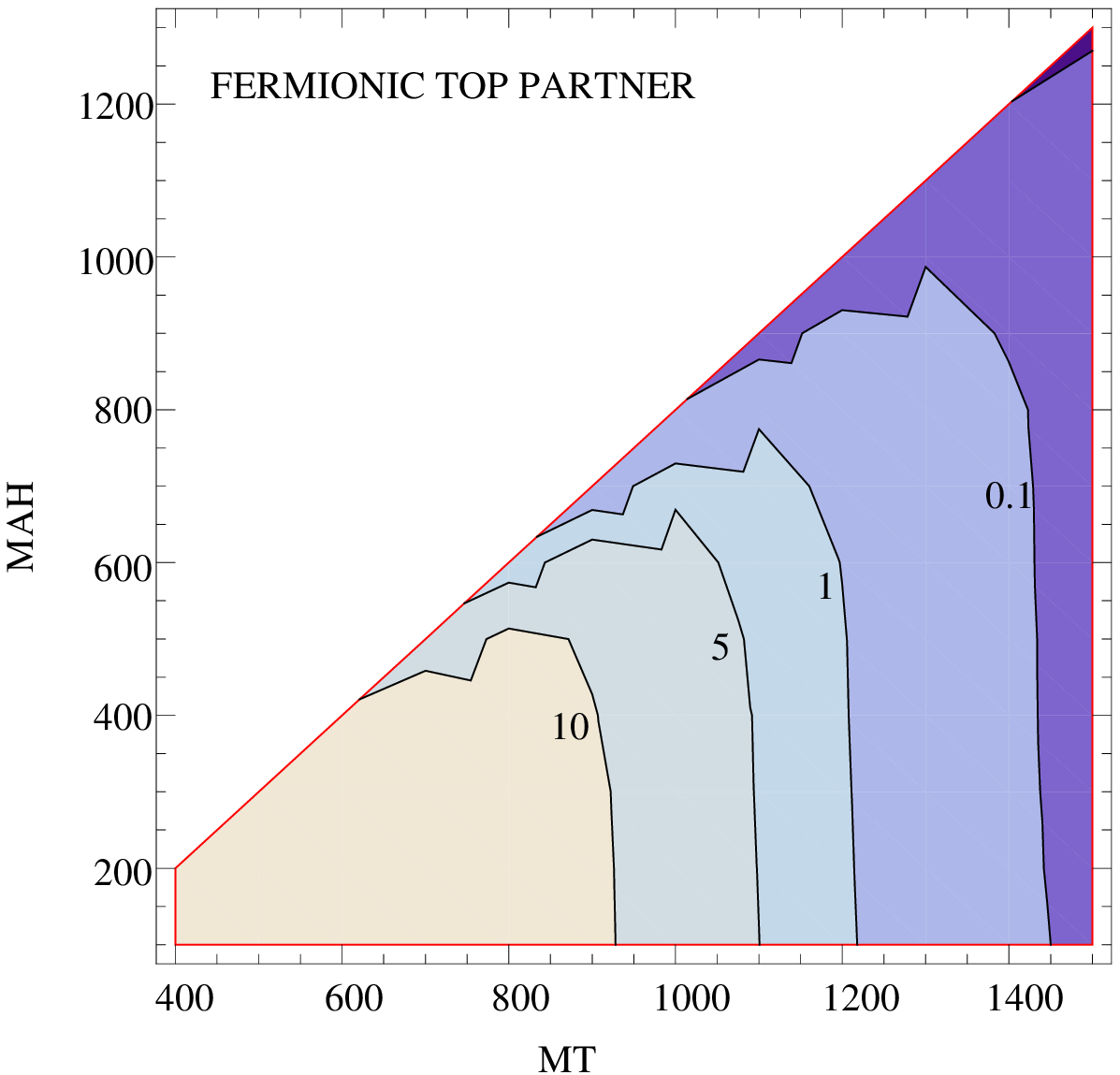,width=8.2cm}
\psfig{file=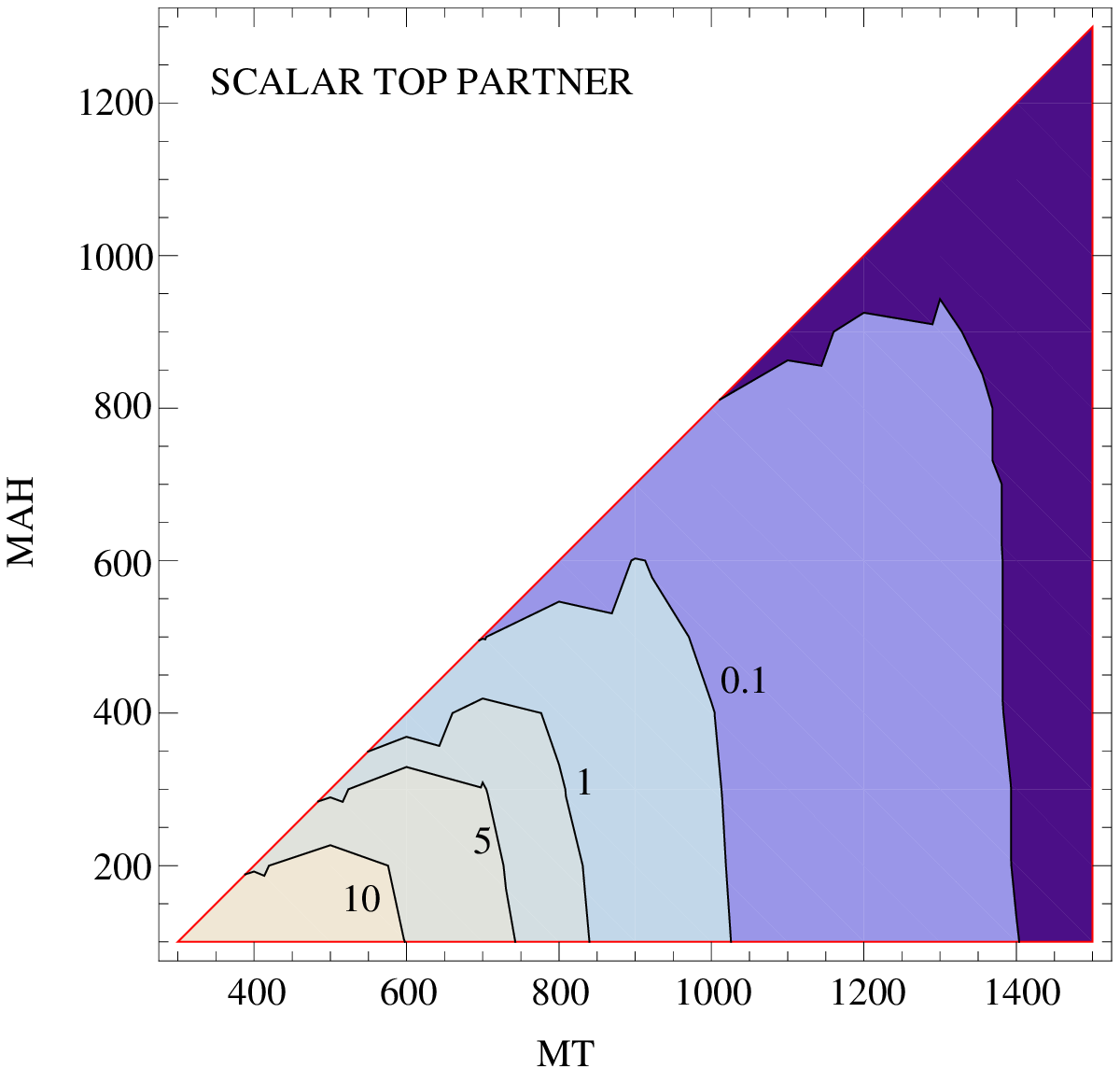,width=8.2cm}}
\caption{Contour in $m_T-m_A$ for $T \to tA$
for the statistical significance at the LHC with an integrated
luminosity of 100 fb$^{-1}$. Left panel is for a fermionic $T$, and the right is 
a scalar $\tilde t$, both decaying to a top plus a missing particle.}
\label{fig:sigT}
\end{figure}

If one of the $t\bar t$ decays hadronically and the other decays leptonically,
the signal may be cleaner. It turns out that if the mass difference $\dm$ is sizable, 
then requiring large missing transverse energy may be sufficient to suppress
the background. However, if $\dm \sim m_t$, then the $\etmiss$ for the signal 
is not much different from the background. On the other hand, the fact that the
$t\bar t$ kinematics can be fully reconstructed in the SM 
implies that the reconstruction
for the signal events would be distinctive due to the large missing mass. Indeed,
the reconstructed $m^r_t$ based on the $\etmiss$ will be far away from the true
$m_t$, and mostly result in an unphysical value. If we impose
\beq
|m_t-m_t^r | >110\  {\gev},
\eeq
we can reach optimal signal identification. The summary plot for 
the statistical significance (the number of $\sigma$) is given in Fig.~\ref{fig:sigT}
at the LHC with an integrated luminosity of 100 fb$^{-1}$,
where the left panel is for a fermionic $T$, and the right is 
a scalar $\tilde t$, both decaying to $t +$ a missing particle.

\subsection{Exotic top signatures}

Searching for exotic events related to the top quark can be rewarding.
First, there exists a variety of natural electroweak models with distinctive 
top partners that should not be overlooked for collider phenomenology.
Second, potentially large couplings of the top quark to new physics 
may result in multiple top quarks from new particle decays.  
Finally, the exotic events have less SM background contamination,
and thus may stand out for discovery even at the early phase of the LHC.
We briefly list a few recent examples.
\begin{itemize}
\item Multiple top quarks and $b$-quarks in the final state 
may help to search for new heavy particles in the electroweak sector
and can be distinctive from the SM backgrounds \cite{Han:2004zh}.
\item Heavy top partners and other KK fermions in the RS model may lead to
unique top-quark and $W$-boson signatures \cite{Contino:2008hi}.
\item New exotic colored states may predominantly couple to heavy quarks
and thus lead to multiple top quarks in the final state \cite{Gerbush:2007fe}.
\item Composite models for the right-handed top-quark may lead to $t\bar t t\bar t$
signals at the LHC \cite{Lillie:2007hd}.
\item Like-sign top quark pairs may indicate  new dynamics \cite{Cao:2004wd}.
\end{itemize}

\section{Summary and Outlook}

The LHC will be a true top-quark factory. With 80 million top-quark pairs plus
34 million single tops produced
annually at the designed high luminosity, the properties of this particle
will be studied to a great accuracy and the deep questions related to the top quark
at the Terascale will be explored to an unprecedented level. Theoretical arguments
indicate that it is highly likely that new physics associated with the top quark
beyond the SM will show up at the LHC. This article only
touches upon the surface of the rich top quark physics, and is focused
on possible new physics beyond the SM in the top-quark sector. The layout of
this article has been largely motivated by experimental signatures for the LHC. 
Interesting signatures covered here include
\begin{itemize}
\item Rare decays of the top quark to new light states, or to SM particles
via the charged and neutral currents through virtual effects of new physics.
\item Top quark pair production via the decay of a new heavy resonance, resulting
in fully reconstructable kinematics for detailed studies.
\item Top quark pair production via the decay of pairly produced top partners,
usually associated with two other missing particles, making the signal
identification and the property studies challenging. 
\item Multiple top quarks, $b$ quarks, and $W^\pm $'s coming from theories
of electroweak symmetry breaking or an extended top-quark sector.
\end{itemize}

The physics associated with top quarks is rich, far-reaching, and exciting. It opens up
golden opportunities for new physics searches, while brings in new challenges as well.
It should be of high priority in the LHC program for both theorists and experimentalists.

\section*{Acknowledgments}
I thank Gordy Kane and Aaron Pierce for inviting me to write on this subject, which
I consider a very important and exciting part of the LHC physics program. 
I would also like to thank Vernon Barger, Tim Tait and 
Lian-Tao Wang for reading and  commenting on the draft. 
This work was supported in part by the US DOE under contract No.~DE-FG02-95ER40896 
and in part by the Wisconsin Alumni Research Foundation.
The work at the KITP was supported by the National
Science Foundation under Grant No. PHY05-51164.

\end{document}